\documentclass[aps,prl,groupedaddress,twocolumn,letterpaper]{revtex4-1}
\input{Qcircuit}

\usepackage[]{graphicx}
\usepackage[]{amsmath}
\usepackage{setspace}
\usepackage{amsthm}
\usepackage{hyperref}
 \usepackage{setspace}

\def\ket#1{| #1 \rangle}
\def\bra#1{\langle #1 |}

\def\Tr{\mathrm{Tr\,}}

\def\exp{\mathrm{exp}}
\def\log{\mathrm{log}}

\def\cO{\mathcal{O}}

\def\cZ{\mathcal{Z}}

\def\eps{\epsilon}

\def\({\left(}
\def\){\right)}

\def\avg#1{\langle #1 \rangle}

\begin{document}

% number in the upper right hand corner of the title page in preprint mode.
%\preprint{}

\title{Preparing thermal states of quantum systems by dimension reduction}

\author{Ersen Bilgin}
\email[]{ersen@caltech.edu}
\affiliation{Institute of Quantum Information, California Institute of Technology, Pasadena, CA, 91125}

\author{Sergio Boixo}
\email[]{boixo@caltech.edu}
\affiliation{Institute of Quantum Information, California Institute of Technology, Pasadena, CA, 91125}

\date{\today}

\begin{abstract}
We present an algorithm that prepares thermal Gibbs states of one
dimensional quantum systems on a quantum computer without any memory overhead, and in a time significantly shorter than other known alternatives. 
Specifically, the time complexity is dominated by the quantity
$N^{\|h\|/ T}$, where $N$ is the size of the system, $\|h\|$ is a
bound on the operator norm of the local terms of the Hamiltonian
(coupling energy), and $T$ is the temperature. Given other results on the complexity
of thermalization, this overall scaling is likely
optimal. For higher dimensions, our algorithm lowers the known
scaling of the time complexity with the dimension of the system by one. 
\end{abstract}

%\keywords{}

\maketitle

Many open problems in condensed matter physics concern strongly
correlated quantum many-body systems.  These are typically not
solvable analytically, and we have to resort to numerical
simulations. Unfortunately, numerical methods tend to fail for general hamiltonians on these systems, due to the exponential scaling of the dimension of the
corresponding Hilbert space.  This problem is one of the main motivations
for the quest of quantum computers. Indeed, quantum computers can
efficiently simulate unitary evolutions of quantum many-body systems
with local
interactions~\cite{feynman_simulating_1982,lloyd_universal_1996},
because they can inherently deal with exponentially large Hilbert
spaces. 

Nevertheless, the preparation of the desired initial state is
still a difficult problem in
general~\cite{kitaev_classical_2002,kempe_complexity_2006,oliveira_complexity_2005,aharonov_power_2009,schuch_computational_2008}.
There have been several proposals to tackle this problem
~\cite{terhal_divincenzo2000,temme_metropolis_2009,cramer2010a,poulin_samplingthermal_2009,Yung2010a}.
Some
significant alternatives have worse complexity
scaling than ours~\cite{terhal_divincenzo2000,poulin_samplingthermal_2009},
while others apply to a restricted set of systems
\cite{cramer2010a,Yung2010a}. The quantum metropolis algorithm~\cite{temme_metropolis_2009}, in particular,
might often be faster, but lacks complexity bounds. The classical
algorithm proposed in~\cite{hastings_solving_2006} can be used to prepare 1D quantum thermal states with only a polynomial time complexity overhead with respect to our method, but its (classical) memory requirements scale exponentially with inverse temperature, and it does not extend to higher dimensional systems.

The time complexity of our method for one dimensional systems is
dominated by the quantity $N^{\|h\|/T}$, where $N$ is the number of
subsystems, $T$ is the temperature, and $\|h\|$ is a bound on the
operator norm of the local terms of the Hamiltonian, the interaction
strength. Note that this scaling is polynomial in $N$. The memory of
the quantum computer scales simply with $N$, an exponential
improvement over general classical algorithms. Our algorithm can also be
massively parallelized, and when run in a cellular automaton
architecture the memory scales as $N^{\|h\|/T}$, but the total time
would be linear in $N$ (the total number of steps would still be the same).  In two and higher dimensions, 
our method lowers the number of effective dimensions by one.  This results in an exponential speedup, but
the exponential scaling with $N$ remains.

The overall scaling appears to be
optimal: the known complexity of thermalizing 1D quantum systems makes
a guaranteed polynomial scaling with temperature highly
unlikely~\cite{aharonov_power_2009,schuch_computational_2008}. We also
expect the grouping of $\|h\|/T$ in the exponent by dimensional
analysis. In other words, the relevant temperature scale is set by the
Hamiltonian. 

It is easier to introduce this method by
explaining the proposal in~\cite{poulin_samplingthermal_2009} first. In order to prepare a thermal state of a given Hamiltonian, the
probability of each eigenstate needs to be set to the correct Gibbs
probability. We can encode the correct probabilities as amplitudes of
a marked state of an ancillary
system.  If a projective
measurement of the ancilla returns the marked state, we succeeded in
preparing the target Gibbs state. The success probability goes like
the probability of projecting a random
state into the target Gibbs
state. If we use the totally mixed state as
the initial state of these projections, the success probability is
$\cZ/d^{N}$, where $d$ is the dimension of each local
subsystem. As a result, the number of trials scales exponentially
with the system size. It is possible to obtain a quadratic speedup over this scaling
using Grover's amplitude
amplification~\cite{poulin_samplingthermal_2009}, but the algorithm still scales
exponentially with system size.

\begin{figure}[htbp]
\begin{center}
\includegraphics[width = 0.8\columnwidth]{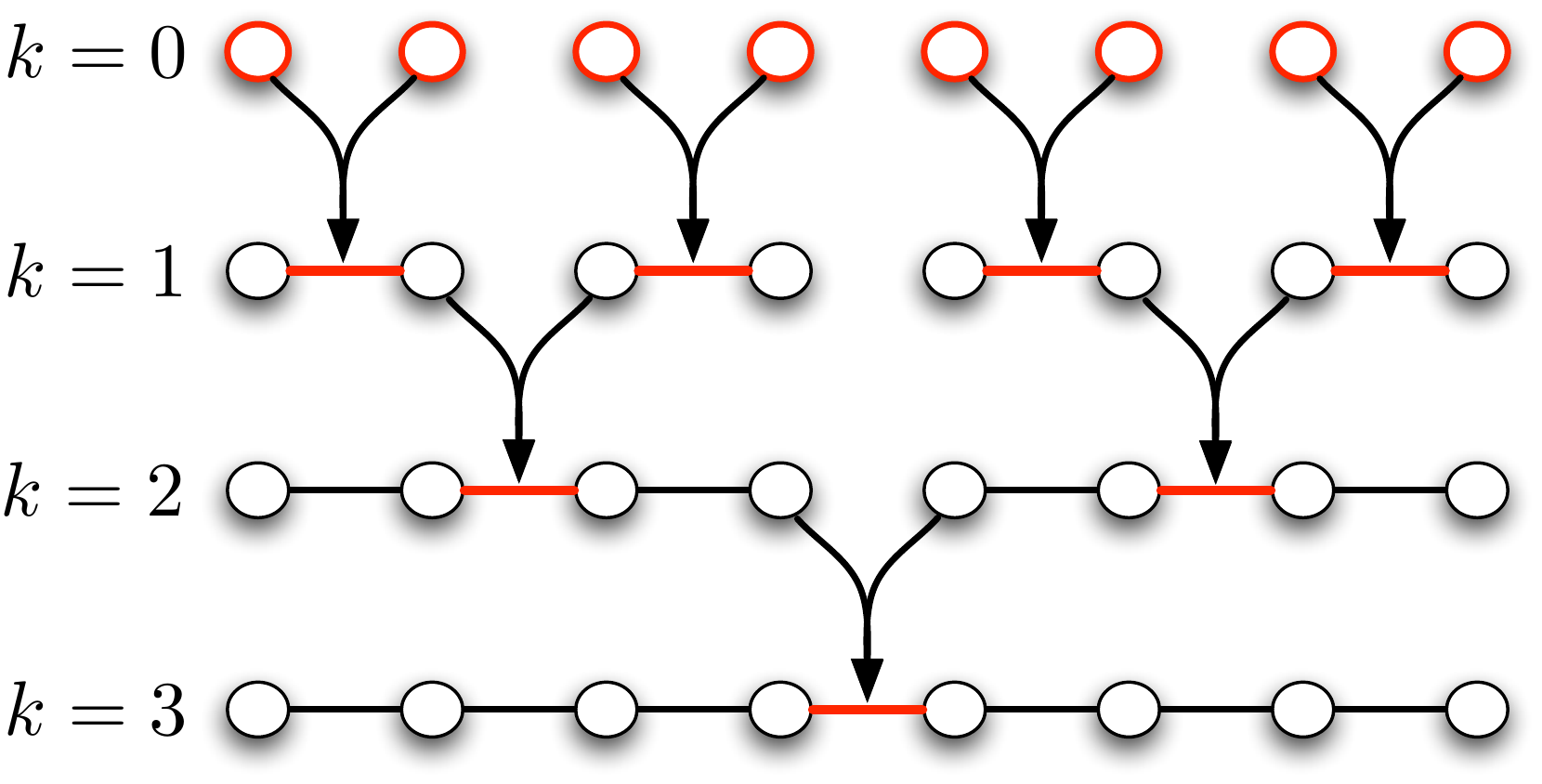}
\caption{The procedure to thermalize an 8-qubit chain.  After thermalizing individual qubits at level $k=0$, we pair them up and merge them by adding the Hamiltonian that connects the two qubits.  This procedure is then repeated recursively as we merge two already thermalized regions of size $2^k$ at level $k$ to obtain a thermalized chain of size $2^{k+1}$ at level $k+1$.}
\label{Fig1}
\end{center}
\end{figure}

We overcome the problem of exponential time cost by dividing
the overall procedure into a sequence of projections and arranging
them so that we only need to rebuild a small
section after most failures (see Fig. \ref{Fig1}). We first
thermalize small regions, which we merge recursively until we
have thermalized the whole system.   Only when
the failures are close to the end of this recursive procedure do we need to
rebuild big sections.  A careful error analysis shows that each of
these merging operations can be implemented with a cost independent of
the system size and the quantum correlation length, resulting in a running time
that is only polynomial in the system size and independent of the
correlation length. This method trivially generalizes
to higher dimensions and reduces the scaling of the cost with the
system dimension by one compared to a direct projection.

We implement the merging perturbatively. Assume that we are given access to copies
of $\rho   \propto e^{-\beta H}$ (from previous steps). The
Hamiltonian $H$ corresponds to the halves
to be merged, but the procedure is more general. We want to generate the state $\rho^{(1)} \propto
e^{-\beta (H+h)}$, where $h$ corresponds to the link between the two
halves. We will see how to generate (with high probability) the state
$\rho^{(\epsilon)} \propto e^{-\beta (H+\epsilon h)}$.  We then repeat the same process to produce the sequence
\begin{align}\label{eq:sequence}
  \rho = \rho^{(0)} \to \rho^{(\epsilon)} \to \rho^{(2\epsilon)} \to
  \cdots \to \rho^{(1)}\;.  
\end{align}
Every transformation in the sequence has some probability of failure,
in which case we restart.   If all of the steps
succeed, we approximate the state $\rho^{(1)}$ with an error of
$\cO(\eps)$. It is important to remark that all the errors in this
paper are in the trace norm. That is, in this case, for input $\rho$, we build a state
$\sigma$ such that $\|\sigma - \rho^{(1)}\|_\Tr \in \cO(\epsilon)$.

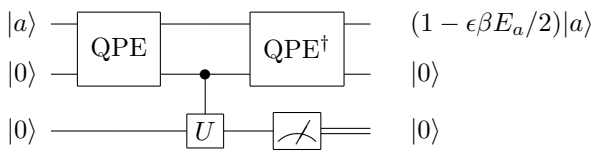
\begin{figure}
 \centering
  \begin{equation*} \!\!\!\!\!\!\!\!\!\!\!\!
  \Qcircuit @C=1em @R=1em {
 \lstick{\ket{a}} &\multigate{1}{{\rm QPE}}  &\qw& \multigate{1}{{\rm QPE^{\dagger}}} &\qw&  \rstick{(1-\eps \beta E_a/2) \ket{a}} \\
 \lstick{\ket{0}} &\ghost{{\rm QPE}} &\ctrl{1} & \ghost{\rm QPE^{\dagger}} &\qw&\rstick{\ket{0}} \\
 \lstick{\ket{0}} &\qw& \gate{U} &\meter&\cw &\rstick{\ket{0}}
 }\quad\quad\quad\quad\quad
 \end{equation*}
 \caption{The conjugation circuit.  First the quantum phase estimation
   circuit (QPE) writes the energy onto an ancilla register.  Then the
   unitary U rotates the second ancilla conditioned on the energy. We
   measure the second ancilla and restart if we do not obtain $\ket 0$.}\label{fig:conj}
\end{figure}

We update the state $\rho \propto e^{-\beta H}$ to $\rho^{(\epsilon)}
\propto e^{-\beta (H+\epsilon h)}$, to first order in $\epsilon$, in
two steps. The first step is probabilistic and updates the
probabilities of the Gibbs state through post-selection. If it fails,
we will be forced to restart, and this will be the dominant part for
the cost of the algorithm. In the second
step we update the eigenbasis to the eigenbasis of $\rho^{(\epsilon)}$.
We now assume perfect phase estimation and perfect dephasing, and
later account for the cost and errors of these operations.

The probabilistic 
transformation of the first step is a conjugation with $e^{-\epsilon \beta
  h/2}$, to first order in $\eps$. We assume that $h\ge 0$. 
We use phase estimation and post-selection as in
Fig. \ref{fig:conj}.  This is similar to the procedure in \cite{poulin_samplingthermal_2009}.  The phase estimation is of the unitary $e^{2\pi i h t}$, with $1/t > h \ge
0$.  It implements the map $ \sum_a P_a\ket{t E_a}\bra 0$, where $P_a$ is a
projection of the system onto the eigenspace of $h$ with energy $E_a$. This energy
gets written in an ancilla register initialized to $\ket 0$. We rotate a second ancilla to
$(1-\epsilon\beta {E}_a/2)| 0 \rangle + \cdots | 1 \rangle$,
conditioned on the value of the previous ancilla register (unitary $U$
in Fig. \ref{fig:conj}), to get the map $\sum_a (1-\epsilon\beta
{E}_a/2) P_a\ket{ t E_a 0} \bra {00}+ \cdots$.  We then undo the phase
estimation, obtaining the map
\begin{align}\label{eq:perfect_conjugation_unitary}
  \sum_a (1-\epsilon\beta {E}_a/2) P_a &\ket {00}\bra{00} +
\cdots \notag \\ &= (1-\epsilon \beta h/2) \ket{00}\bra{00} + \cdots 
\end{align}
Finally we
measure the ancilla, and fail unless we obtain $\ket 0$.  
We denote the result obtained by applying the above map to $\rho$ as
\begin{align}\label{eq:conjmap}
  \rho_{\rm prob} \propto (1-\epsilon \beta h/2)\rho (1-\epsilon \beta
  h/2)\;.
\end{align}
The success probability is
\begin{align}\label{eq:probability}
  p \ge 1-\epsilon \beta \| h\| \;.
\end{align}

It can be seen that the conjugation with $1-\epsilon \beta h/2$ of the
previous paragraph updates all
the probabilities of the eigenstates correctly to first order in $\epsilon$. It also
implements the appropriate transformation for the degenerate subspaces
of $\rho$. Next, we update the eigenbasis from $H$ to $H + \epsilon
h$, using the adiabatic approximation. This approximation
can be seen as a consequence of the Zeno effect, which can be
achieved through measurements or dephasing~\cite{childs_quantum_2002,boixo_quantum_2009}. Here we
implement the Zeno effect directly.  More specifically, we use pure dephasing in the eigenbasis of $H + \epsilon h$
(see~\cite{boixo_quantum_2009}). 

Denote by $\{ P_{k^\epsilon}\}$ the projectors on the eigenbasis of
$H+\epsilon h$. For input $\rho$, the operation $\sum_{k^\epsilon}
P_{k^\epsilon} \rho P_{k^\epsilon}$ is pure dephasing on this
eigenbasis.  To qualify the effect of dephasing on $\rho_{\rm prob}$,
we start by writing the result of this ideal operation as~\cite{boixo_quantum_2009}
   \begin{align}\label{eq:dephasing}
     \sum_{k^\epsilon} P_{k^\epsilon} \rho_{\rm prob} P_{k^\epsilon} =
     \lim_{\sigma \to \infty} \int dt\,
     g_\sigma(t) U_\epsilon(t) \rho_{\rm prob}  U_\epsilon^\dagger(t)\;,    
   \end{align} 
   where $g_\sigma$ is the density function of a Gaussian distribution
   with $0$ mean and standard deviation $\sigma$, and $U_\epsilon(t) =
   e^{-i (H + \epsilon h) t}$.  This equality is easy to check
   directly. We now use a Dyson series to second order for the time evolution:
   \begin{align}\label{eq:dyson_series_time_evolution}
     U_\epsilon(t) &= U_0(t) - i \epsilon A(t) - \epsilon^2 B(t)\\
   U_0(t) &= e^{-i H t} \\ 
   A(t) &= \int_0^t dt_1\,U_0(t-t_1) h U_0(t_1) \\
   B(t) &= \int_0^t dt_1\,\int_0^{t_1} dt_2\, U_0(t-t_1) h
   U_0(t_1-t_2) h U_\epsilon(t_2)\;. \nonumber         
   \end{align}
We then insert  terms to first order in $\epsilon$ from
Eq.~\eqref{eq:dyson_series_time_evolution} into Eq.~\eqref{eq:dephasing}. The second (and higher, after expanding further) order terms can all be bounded by  $\cO(\epsilon^2 \|h\|^2)$ in the trace-norm.

We want to compare the state obtained after the dephasing to $\rho^{(\epsilon)}$, to first
order in $\epsilon$.  With this goal, we use the Dyson series in imaginary time to
expand $\rho^{(\epsilon)}$.  In this expansion we need intermediate states $\rho(\tilde\beta) = e^{-\tilde\beta
  H}/\cZ$, where the partition function is always at inverse
temperature $\beta$, $\cZ = \Tr e^{-\beta H}$. We obtain, to
second order,
\begin{align}\label{eq:dyson_imaginary_time}
  \rho^{(\epsilon)}&=  \rho - \epsilon \int_0^\beta d\beta_1 \,\rho(\beta-\beta_1) h
  \rho(\beta_1) \\ 
  +&\epsilon^2 \int_0^\beta\! d  \beta_1\int_0^{\beta_1}d
  \beta_2 \,\rho(\beta - \beta_1) h \rho(\beta_1-\beta_2)h \rho(\beta_2)\nonumber.
\end{align}
The second and higher order corrections can be bounded by
$\cO(\eps^2\beta^2\|h\|^2)$ in trace-norm.  We assume that $\beta \ge
1$, so this error dominates the error $\cO(\epsilon^2 \|h\|^2)$ above.

We aim to obtain the first
order expression in Eq.~\eqref{eq:dyson_imaginary_time} with our procedure.
The spectral decomposition of $\rho$ is $ \sum_k p_k P_k$, where $P_k$
is the projector to the subspace spanned by eigenstates with
eigenvalue $E_k$ of $H$, and $p_k = e^{-\beta E_k}/\cZ$.  The term
proportional to $\epsilon$ in the expansion of $\rho^{(\epsilon)}$ in Eq.~\eqref{eq:dyson_imaginary_time} can be rewritten in terms of these projectors, up to normalization, as:
\begin{align}
  -\epsilon \sum_k p_k \(P_k h  P_k + \sum_{l\ne k} \frac {P_l h P_k +
  P_k h P_l} {E_l - E_k} \). \label{eq:foa}
\end{align}
Finally, using $H = \sum_k E_k P_k$ in Eq.~\eqref{eq:dephasing}, with
the expansion~\eqref{eq:dyson_series_time_evolution}, gives the first order correction Eq.~\eqref{eq:foa}.  

Until now, we were considering the behavior of our
 algorithm assuming perfect phase estimation and dephasing, which is not
 realistic. We now account
 for the effects of the errors inherent in the two parts of our
 perturbative Hamiltonian update algorithm when implemented using only
 dynamical evolution with the Hamiltonian $H+k h$ for $0\le k\le1$ . We quantify
 the total cost in terms of the evolution time.
 
 For the conjugation circuit, we use high precision phase
 estimation~\cite{knill_optimal_2007,somma_quantum_2008,poulin_samplingthermal_2009,chiang_quantum_2010}. The
 cost (evolution time with $h$), for accuracy $\delta$ and error
 $\varepsilon$, scales as $\cO(\log(1/\varepsilon)/\delta)$. This implements the transformation
   \begin{align}
     \sum_a P_a \otimes \Bigg(  \Bigg( \sum_\pm c_a^\pm
     \ket{t E_a^\pm} + q_a \ket {\xi_a}\Bigg)   \bra 0 \Bigg)\;,
   \end{align}
   with $t |E_a - E_a^\pm| \le \delta$ and $q_a \le \varepsilon$.  The
   $t$ is the evolution time of the unitary $e^{2 \pi i h t}$ used for
   the phase estimation, with $1/t \ge h \ge 0$ as above.  The term
    $\ket{\xi_a}$ groups all the states of the ancilla register with a
    reading outside the goal accuracy $\delta$.%, and its final effect
%    can be bounded by $\varepsilon$ in the trace norm because they are
%    orthogonal to $\ket{t E_a ^\pm}$. 
   
   First consider the effect of the error term $\sum_a q_a P_a
   \ket{\xi_a}$. All the manipulations conserve the projectors $P_a$
   and do not increase the norm of $\ket{\xi_a}$.   Therefore, the final error
   due to these terms, on input $\sigma$, can be bounded with terms
   like $\| \sum_a q_a P_a \sigma \|_{\rm tr} \le \varepsilon$. The
   final error can increase as a result of the normalization when
   projecting onto the post-selected state, if the preparation is
   successful. We will see that this effect is negligible. We then
   rotate a second ancilla and undo the phase estimation to obtain an
   approximate version of the conjugation unitary in
   Eq.~\eqref{eq:perfect_conjugation_unitary}. The undo of the phase
   estimation incurs in an error bounded by~$\cO(\epsilon \beta
   \delta)$ in the trace norm, and the rest of  the errors are $\cO(\varepsilon)$ as before. If we choose $\varepsilon$ and $\epsilon \beta \delta$ to be $
\cO(\epsilon^2 \beta^2 \|h\|^2)$, we can implement the map in
Eq.~\eqref{eq:conjmap} with the same probability~\eqref{eq:probability} as
before, trace norm error $\cO(\epsilon^2\beta^2 \|h\|^2)$,
and cost (evolution time)
   \begin{align}
     \cO\(\log(1/(\epsilon\beta\|h\|))/(\epsilon \beta\|h\|^{2})\).
   \end{align}

Finally, we deal with the errors related to imperfect dephasing.
Irrespective of the implementation of dephasing, imperfect accuracy in
the dephasing results in an operation where  only phases between
eigenstates with relative gap bigger
than some bound $\zeta$ are suppressed. This is modeled by the
operation 
  \begin{align}
    \sigma \to \sum_{j^\epsilon,k^\epsilon : |E^\epsilon_k - E^\epsilon_j| \le
    \zeta}   P_{k^\epsilon} \sigma P_{j^\epsilon}
  \end{align}
(cf. Eq.~\eqref{eq:dephasing}).
 %  The difference with the perfect dephasing operation are the terms
%   with $ 0 <  |E^\epsilon_k - E^\epsilon_j| \le \zeta$. 
We 
  obtain imperfect dephasing if we choose the standard deviation of the Gaussian in
  Eq.~\eqref{eq:dephasing} to be $\sigma = \cO(1/\zeta)$, instead of (approaching)
  infinity~\footnote{We can truncate the tails of the Gaussian distribution to
  ensure a worst time cost of the order of the average cost, and the
  resulting error is already dominated by the imperfect dephasing.}.

We analyze the effect of imperfect dephasing errors by the following purely formal
mathematical procedure. For a given Hamiltonian $H+\epsilon h$, we
define the Hamiltonian $\tilde H$ which is similar to $H+\epsilon h$,
but such that all the gaps are at least $\zeta$. We do this by
increasing the degeneracy: we group all eigenvalues
in bins with relative gap  $\zeta$. The imperfect
dephasing (or accuracy $\zeta$) would be fundamentally
exact if the target Hamiltonian
was $\tilde H$, because all the gaps would be bigger than the accuracy. Let $\chi = \tilde H - (H
+\epsilon h)$, with $\|\chi\| \le \zeta$ in operator
norm. 

By an expansion similar to that of
Eq.~\eqref{eq:dyson_imaginary_time} we
can write $\rho^{(\eps)} = \tilde{\rho}^{(\eps)} + \cO(\beta
\zeta)$, where $\tilde{\rho}^{(\eps)} = e^{-\beta\tilde
  H}/\Tr e^{-\beta \tilde H}$ and the bound is, as always, in the
trace norm~\footnote{An analogous bound holds between $\rho$ and the
  Gibbs state of $\tilde H - \epsilon h$. We also need a bound on the
  difference between the partition functions of $\rho^{(\eps)}$ and
  $\tilde{\rho}^{(\eps)} $, which can be found in~\cite{poulin_samplingthermal_2009}}.  Similarly, we can write
$U_{\eps}(t)$ in terms of $\tilde{U}_\eps(t) = e^{-i
  \tilde{H} t}$ and $\chi$ using Dyson series.  Plugging these
expressions into Eq.~\eqref{eq:dephasing} with $\sigma = c/\zeta$ for
a constant $c = \cO(\log (1/\varepsilon))$, we get
\begin{align}
      \int dt\,
     g_{c/\zeta}(t) U_\epsilon(t) \rho_{\rm prob}
     U_\epsilon^\dagger(t) = \rho^{(\eps)} + \cO(\beta \zeta)\;, 
\end{align} 
where $g_{c/\zeta}$ is a Gaussian with standard deviation $c/\zeta$.
   Now choosing $\zeta = \eps^2 \beta \|h\|^2$ we see that we can
   transform $\rho$ to $\rho^{(\eps)}$ with the probability of
   Eq.~\eqref{eq:probability}, trace norm error $\cO(\eps^2 \beta^2 \|h\|^2)$, and cost (evolution time)
   \begin{align}\label{eq:step_evolution_time}
      \cO(\log(1/(\eps\beta \|h\|))/(\eps^2 \beta \|h\|^2))\;.    
   \end{align}

We can merge two regions already thermalized into one large thermal
region with the two subroutines just described using a sequence of
small perturbative steps (see Eq.~\eqref{eq:sequence}). Each step is
successful with the probability given in
Eq.~\eqref{eq:probability}. The average number of steps until we
generate a complete sequence without failures is $\avg{m} \in
\cO(e^{\beta \|h\|})$~\footnote{This is also known from the theory of
  success runs. We give the average cost, but the tail has an exponential decay rate, so
  the worst case cost is similar (see, for
  instance~\cite{balakrishnan_runs_2001}).}. Each time that we fail we
need to produce two new thermal regions to be merged. The average
number of failures is $\avg{\alpha} \in \cO(e^{\beta \|h\|})$. 
  
Now we analyze the average number of steps
$\avg{\tau(k)}$ required to prepare a thermalized chain of
length $2^k$ at level $k$ (see Fig. \ref{Fig1}). Since $\alpha$ and
$\tau(k-1)$ are independent random variables, the
expectation value of $\tau(k)$ is
\begin{align}
  \langle \tau(k) \rangle = 2 \langle
\alpha \rangle \langle \tau(k-1) \rangle + \langle m \rangle\;. 
\end{align} 
This gives $\avg{\tau(\log N)} \in \cO( \exp((\beta\|h\|+\log 2)\log
N))$. Similarly, the total error is $\cO(N \epsilon
\beta^2\|h\|^2)$. If we choose $\eps$ = $\bar{\eps}/(N \beta^2
\|h\|^2)$, we get a total error of $\cO(\bar{\eps})$ in
trace-norm. Finally, using Eq.~\eqref{eq:step_evolution_time} for the
evolution time of each step, we obtain the dominant contribution to
the total evolution time $\beta N^{\beta \|h\|} /\bar \epsilon^2$.

We have presented an algorithm that prepares a thermal state of a 1D
quantum system in time polynomial in the system size and exponential
in the inverse temperature (as required by the existence of
QMA-complete ground state problems in 1D).  This algorithm can be
generalized into $D$ dimensions. At level $k$ of the recursion, we
have built squares (for 2D) or cubes that are now merged. We do
not get polynomial scaling with system size for $D>1$ because the
intersection of two neighboring regions goes like
$N^{D-1}$. Note
that this is to be expected because there exist 2D ground states with
constant gap that encode the solution to 
NP-complete problems. A careful analysis confirms that the
time complexity is dominated by the operations at the top level, and the
dominating factor is $\beta e^{2 \beta \|h\| D N^{D-1}}/\bar{\eps}^2$. This is an exponential speedup
from the known $\exp(O(N^D))$. The memory requeriments still scale
with the number of sites of the model, $N^D$.

There are also several possible improvements to the scaling of this
algorithm.  If one is interested in thermalizing a classical system
with a small quantum perturbation one can first solve for the
classical part of the Hamiltonian.  Then, one would only need to use
projections for the quantum perturbation. Also, if one is interested
in thermalizing a quantum system with short-ranged quantum
correlations, one can also use belief
propagation~\cite{hastings_quantum_2007,leifer_quantum_2008,poulin_belief_2008,bilgin_coarse-grained_2010}
to reduce the storage requirements from $\cO(N)$ qubits to $\cO(l\,
\log(N))$, where $l$ is a constant related to the quantum correlation
length. This can be done by tracing out parts of the blocks that do
not share any entanglement with the boundary to be merged.    The time complexity of the algorithm remains the same as before. Note that the cost (both in memory and time) of the classical algorithm of quantum belief
propagation for 1D quantum systems is exponential in $l \sim
1/T$. This bound is only heuristic, and similar to~\cite{hastings_solving_2006}.

\begin{acknowledgments}
We acknowledge helpful discussions with D. Poulin. This work is supported by DoE under Grant No. DE-FG03-92-ER40701, and NSF under Grant No. PHY- 0803371.
\end{acknowledgments}

\bibliography{BB10a}

\end{document}